\documentclass[aps,twocolumn,showpacs,byrevtex,prl,reprint]{revtex4-1}
\usepackage{epsfig}
\usepackage{graphicx}
\usepackage{dcolumn}
\usepackage{bm}
\usepackage{overpic}
\usepackage{subfigure}
\usepackage{float}
\usepackage{color}
\usepackage{amsmath}
\usepackage{mathcomp}
\usepackage{mathrsfs}
\usepackage{multirow}
\usepackage{rotating}

\begin{document}
\normalsize
\parskip=5pt plus 1pt minus 1pt

\title{ \boldmath Observation of the $W$-Annihilation Decay $D^{+}_{s} \rightarrow \omega \pi^{+}$ 
and Evidence for $D^{+}_{s} \rightarrow \omega K^{+}$ }
\vspace{-1cm}

\author{
   \begin{small}
    \begin{center}
      M.~Ablikim$^{1}$, M.~N.~Achasov$^{10,d}$, S. ~Ahmed$^{15}$, M.~Albrecht$^{4}$, M.~Alekseev$^{56A,56C}$, A.~Amoroso$^{56A,56C}$, F.~F.~An$^{1}$, Q.~An$^{53,43}$, J.~Z.~Bai$^{1}$, Y.~Bai$^{42}$
% ...
         \\
         \vspace{0.2cm}
   (BESIII Collaboration)\\
\vspace{0.2cm} {\it
$^{1}$ Institute of High Energy Physics, Beijing 100049, People's Republic of China\\
% ...
$^{4}$ Bochum Ruhr-University, D-44780 Bochum, Germany\\
% ...
$^{10}$ G.I. Budker Institute of Nuclear Physics SB RAS (BINP), Novosibirsk 630090, Russia\\
% ...
$^{15}$ Helmholtz Institute Mainz, Johann-Joachim-Becher-Weg 45, D-55099 Mainz, Germany\\
% ...
$^{42}$ Southeast University, Nanjing 211100, People's Republic of China\\
$^{43}$ State Key Laboratory of Particle Detection and Electronics, Beijing 100049, Hefei 230026, People's Republic of China\\
% ...
$^{53}$ University of Science and Technology of China, Hefei 230026, People's Republic of China\\
% ...
$^{56}$ (A)University of Turin, I-10125, Turin, Italy; (B)University of Eastern Piedmont, I-15121, Alessandria, Italy; (C)INFN, I-10125, Turin, Italy\\
% ...
$^{d}$ Also at the Novosibirsk State University, Novosibirsk, 630090, Russia\\
% ...
}\end{center}
\vspace{0.4cm}
\end{small}
}

\affiliation{}
\vspace{-4cm}
\date{\today}

\begin{abstract}
We report on the observation of the $W$-annihilation decay $D^{+}_{s} \rightarrow \omega \pi^{+}$ 
and the evidence for $D_{s}^{+} \rightarrow \omega K^{+}$ with a data sample 
corresponding to an integrated luminosity of 3.19 ${\mbox{\,fb}^{-1}}$ 
collected with the BESIII detector at the 
center-of-mass energy $\sqrt{s} = 4.178$ GeV.  
We obtain the branching fractions 
$\mathcal{B}(D^{+}_{s} \rightarrow \omega \pi^{+}) 
= (1.77\pm0.32_{{\rm stat.}}\pm0.11_{{\rm sys.}}) \times 10^{-3}$ 
and $\mathcal{B}(D^{+}_{s} \rightarrow \omega K^{+}) 
= (0.87\pm0.24_{{\rm stat.}}\pm0.07_{{\rm sys.}}) \times 10^{-3}$,
respectively. 
\end{abstract}
\maketitle
Within the Standard Model of particle physics, direct CP violation (CPV) 
in hadronic decays can only be induced in decays that proceed via at least two distinct decay amplitudes with 
non-trivial strong and weak phase differences~\cite{Cheng:2012wr,Li:2012cfa,Beast1}. 
In the charm sector, examples for such decays are the 
single Cabibbo suppressed (SCS) decays involving $W$-annihilation process~\cite{Cheng:2012wr,Li:2012cfa,Beast1,HaiYangCheng1}, 
for example $D^{+}_{s} \rightarrow \omega K^{+}$,  
and other $VP$ final state ($V$ and $P$ refer to the vector and pseudo-scalar mesons, respectively).
However, in $D$ decays, the $W$-annihilation amplitude is dominated by the nonfactorizable long-distance amplitude induced by the final-state interaction (FSI).  
The corresponding theoretical calculation is very unreliable and  
results in some ambiguity in the prediction of the branching fractions (BFs) and the CPV of related decays. 
Instead, the experimental BF measurement of decays through the $W$-annihilation is used as an 
input in theoretical calculations~\cite{Li:2012cfa,Beast1,HaiYangCheng1,HaiYangCheng2}.
Therefore, the BF of the Cabibbo favored (CF) decay $D^{+}_{s} \rightarrow \omega \pi^{+}$
proceeding only via the $W$-annihilation process provides direct knowledge of the 
$W$-annihilation amplitude.

Compared with the SCS decays, the BF of $D^{+}_{s} \rightarrow \omega \pi^{+}$ 
is expected to be larger and may be measured with a higher precision, is thus 
a more useful experimental input in the $W$-annihilation amplitude determination. 
Evidence for this decay was first reported by CLEO in 1997 and a ratio 
$\frac{\Gamma(D^{+}_{s} \rightarrow \omega \pi^{+})}{\Gamma(D^{+}_{s} \rightarrow \eta \pi^{+})}
= 0.16\pm0.04\pm0.03$ was measured based on 4.7 fb$^{-1}$ 
data taken at the $\Upsilon(4S)$ peak~\cite{CLEOobserve}.  
Later in 2009, using a data sample corresponding to an integrated luminosity of  
0.586 fb$^{-1}$ taken at center of mass energy $\sqrt{s} = 4.170~{\rm GeV}$, 
CLEO observed $6.0\pm2.4$ signal events and measured 
the absolute BF to be $(2.1\pm0.9)\times 10^{-3}$~\cite{CLEOcomega}. 

With the previously measured $W$-annihilation $D \rightarrow VP$ decay as input, 
for $D^{+}_{s} \rightarrow \omega K^{+}$, when considering the $\rho-\omega$ mixing, 
the BF and CPV are predicted to be $0.07\times10^{-3}$ and $-2.3\times10^{-3}$~\cite{Beast1}, respectively, 
where this predicted CPV is in the largest order of magnitude in charmed meson decays. 
However, the corresponding values are predicted to be 
$0.6\times10^{-3}$ and $-0.6 \times 10^{-3}$~\cite{Beast1}, respectively, 
when the $\rho-\omega$ mixing is negligible. 
Therefore, the search for $D^{+}_{s} \rightarrow \omega K^{+}$
can also test the $\rho-\omega$ mixing effect 
and if $D^{+}_{s} \rightarrow \omega K^{+}$ is a good decay mode to search for CPV in charm decays. 

In this Letter, we report on measurements of the absolute BFs of the 
hadronic decays $D_{s}^{+}\rightarrow \omega \pi^{+}$ and 
$D_{s}^{+}\rightarrow \omega K^{+}$.
Charge conjugation is implied throughout this Letter unless explicitly stated.
At the center-of-mass energy of $\sqrt{s} = 4.178$~GeV, the $D_{s}^{+}$ meson is predominantly produced through the process 
$e^+e^- \to D_{s}^{*+}D_{s}^{-} + c.c.$, where the 
$D^{*+}_{s}$ decays to either $\gamma D_{s}^{+}$ or $\pi^{0} D_{s}^{+}$. 
As a consequence, any event that contains a $D^{+}_{s}$ meson also contains a $D^{-}_{s}$ meson.
This condition enables the usage of a powerful ``double tag (DT)" technique~\cite{tagmethod} to measure absolute BFs. 
Events with at least one $D_{s}^{-}$ candidate reconstructed, 
which are referred to as ``single tag (ST)" events,  
provide a sample with a known number of $D^{+}_{s}D^{-}_{s}$ pairs. 
The absolute BF of the signal mode ($\mathcal{B}_{{\rm sig}}$) is determined by forming 
DT signal events from the tracks and clusters not used to 
reconstruct the $D_{s}^{-}$ candidate. The value of the BF
can then be obtained using
\begin{eqnarray}
\begin{aligned}
\label{Eq:Br}
\mathcal{B}_{{\rm sig}} = \frac{Y_{{\rm sig}}}{\sum_{i}{\frac{Y_{{\rm tag}}^{i}\epsilon_{{\rm tag,sig}}^{i}}{\epsilon_{{\rm tag}}^{i}}}}, 
\end{aligned}
\end{eqnarray}
where $Y_{{\rm sig}}$ is the DT signal yield, 
$\epsilon^{i}_{{\rm tag,sig}}$ is the DT efficiency,
and $Y_{{\rm tag}}^{i}$ and $\epsilon_{{\rm tag}}^{i}$ 
are the ST yield and the ST efficiency of the $i$th tag mode, respectively. 

A detailed description of the BESIII detector can be found in Ref.~\cite{detector}. 
Two endcap time-of-flight systems were upgraded with multi-gap resistive plate chambers~\cite{MRPC}.
Monte Carlo (MC) simulations of BESIII detector are based on {\sc geant4}~\cite{sim}. 
The generic MC sample includes all 
known open charm processes, 
initial state radiation to the $J/\psi$ or the $\psi(3686)$, 
and the continuum process. 
The open charm processes are generated with {\sc conexc}~\cite{Ping:2013jka}, considering the effects from 
initial state radiation and final state radiation.  
The decay modes with known BFs are simulated with {\sc evtgen}~\cite{EvtGen}.
The generators {\sc kkmc}~\cite{KKMC} 
and {\sc babayaga}~\cite{BABAYAGA} are used to simulate the continuum.
The generic MC sample, corresponding to an effective luminosity of $110.6 {\mbox{\,fb}^{-1}}$,  
is used to determine the ST efficiency and estimate the background. 
The signal MC sample, two million of events with one $D_{s}$ meson decaying to the signal modes
and the other one decaying to anything, as used in generic MC, 
is generated to estimate the DT efficiency. 

ST and DT candidates are constructed from individual $\pi^{+}$, $K^{+}$, $K^{0}_{\rm S}(\pi^{+}\pi^{-})$ and 
$\pi^{0}/\eta(\gamma\gamma)$ candidates in an event. 
All charged tracks, except the $K_{\rm S}^{0}$ daughters, 
are required to originate from within 10~cm (1~cm) 
along (perpendicular to) the beam axis with respect to the interaction point (IP). 
Their polar angles ($\theta$) are required to be within $|\cos \theta|<0.93$. 
The combination of information about the energy loss in the multi-layer drift chamber 
and the time-of-flight is used to identify the species of charged particles by
calculating a probability $P(K^{+})[P(\pi^{+})])$ 
that the track satisfies the hypothesis of being a $K^{+}(\pi^{+})$. 
The $K^{+}$ and $\pi^{+}$ candidates are required to satisfy $P(K^{+}) > P(\pi^{+})$ and $P(\pi^{+}) > P(K^{+})$, respectively.

For $K_{\rm S}^{0}$ candidates, the combination of 
two oppositely charged tracks whose mass hypotheses are set to the pion, 
without particle identification (PID) applied and with distances of 
closest approach to the IP less than 20~cm along the 
beam axis, is required to have an invariant mass in the interval $[0.487,\,0.511]~$GeV$/c^{2}$. 

The energy of each photon from the $\pi^{0}/\eta$ is required to be 
larger than 25~(50)~MeV in the barrel (endcap) region of the 
electromagnetic calorimeter~\cite{detector}. The opening angles 
between the shower and all the charged tracks should be larger than $10^{\circ}$. The invariant mass of the 
$\gamma\gamma$ pair is required to be within 
the asymmetric intervals $[0.115,\,0.150]$ and $[0.490,\,0.580]$ GeV$/c^{2}$ for $\pi^{0}$ and $\eta$. 
Furthermore, the $\pi^{0}/\eta$ 
candidates are constrained to their nominal mass~\cite{PDG} 
via a kinematic fit to improve their energy and momentum resolution. 

The momenta of all pions, except those from the decay of $K^{0}_{\rm S}$, 
are required to be greater than 0.1~GeV/$c$ in order to reject low momentum pions produced in the $D^{*}$ decay.
The ST events are selected by reconstructing a $D_{s}^{-}$ meson in the two highest purity 
decay modes $D_{s}^{-} \rightarrow K^{0}_{\rm S} K^{-}$ and $K^{+}K^{-}\pi^{-}$. 
For the $D_{s}$ candidate, a recoil mass $M_{{\rm rec}}$ is defined as
$M_{{\rm rec}} = \sqrt{(E_{{\rm tot}}-\sqrt{p_{D_{s}}^{2}+m_{D_{s}}^{2}})^{2} 
        - |\vec{p}_{{\rm tot}} - \vec{p}_{D_{s}}|^{2}}$,
where $E_{{\rm tot}}$, $p_{D_{s}}$, $m_{D_{s}}$, $\vec p_{{\rm tot}}$ and $\vec{p}_{D_{s}}$
are the total energy of $e^{+}e^{-}$, the momentum of the $D_{s}$ candidate, the nominal mass of $D_{s}$~\cite{PDG}, 
the three-momentum vector of the colliding $e^{+}e^{-}$ system, 
and the three-momentum vector of the reconstructed $D_{s}$ candidate,
respectively. 
If the selected $D_s^{-}$ candidate originates directly from the $e^{+}e^{-}$ annihilation,
$M_{{\rm rec}}$ peaks at the $D_{s}^{*}$ mass ($m_{D_{s}^{*}}$).
The other $D_{s}$ candidate has a broader distribution around $m_{D_{s}^{*}}$ in the $M_{{\rm rec}}$ spectrum.
The $D_{s}$ invariant mass and $M_{{\rm rec}}$ of the candidate are required to fall into the ranges 
$[1.90,\,2.03]~$GeV$/c^{2}$ and $[2.05,\,2.18]~$GeV$/c^{2}$, respectively.
If there are multiple candidates in an event, the one with $M_{{\rm rec}}$ 
closest to $m_{D_{s}^{*}}$~\cite{PDG} is chosen. 

\begin{figure}[htp]
\begin{center}
\begin{minipage}[b]{0.235\textwidth}
\epsfig{width=0.98\textwidth,clip=true,file=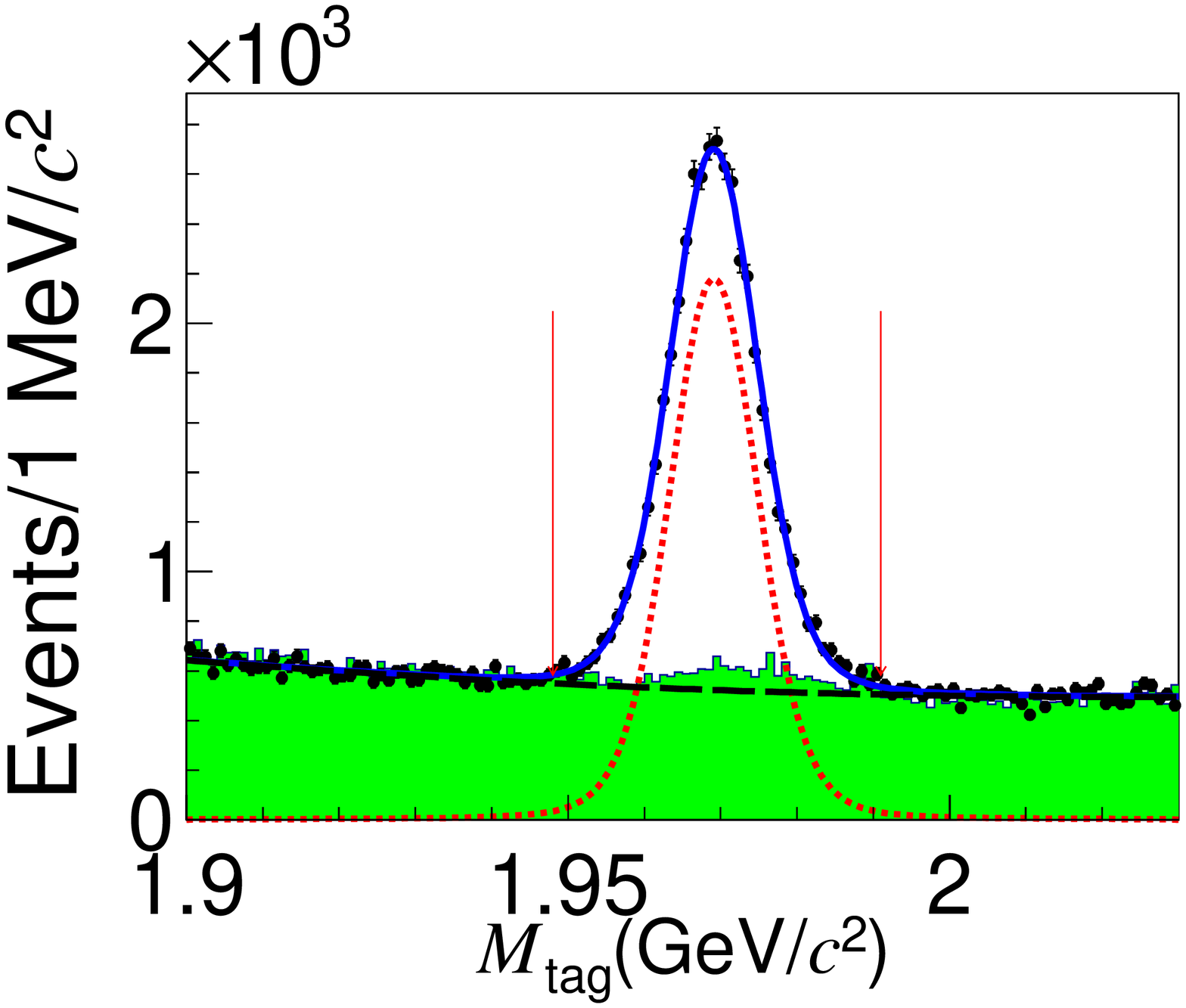}
\put(-20,65){(a)}
\end{minipage}
\begin{minipage}[b]{0.235\textwidth}
\epsfig{width=0.98\textwidth,clip=true,file=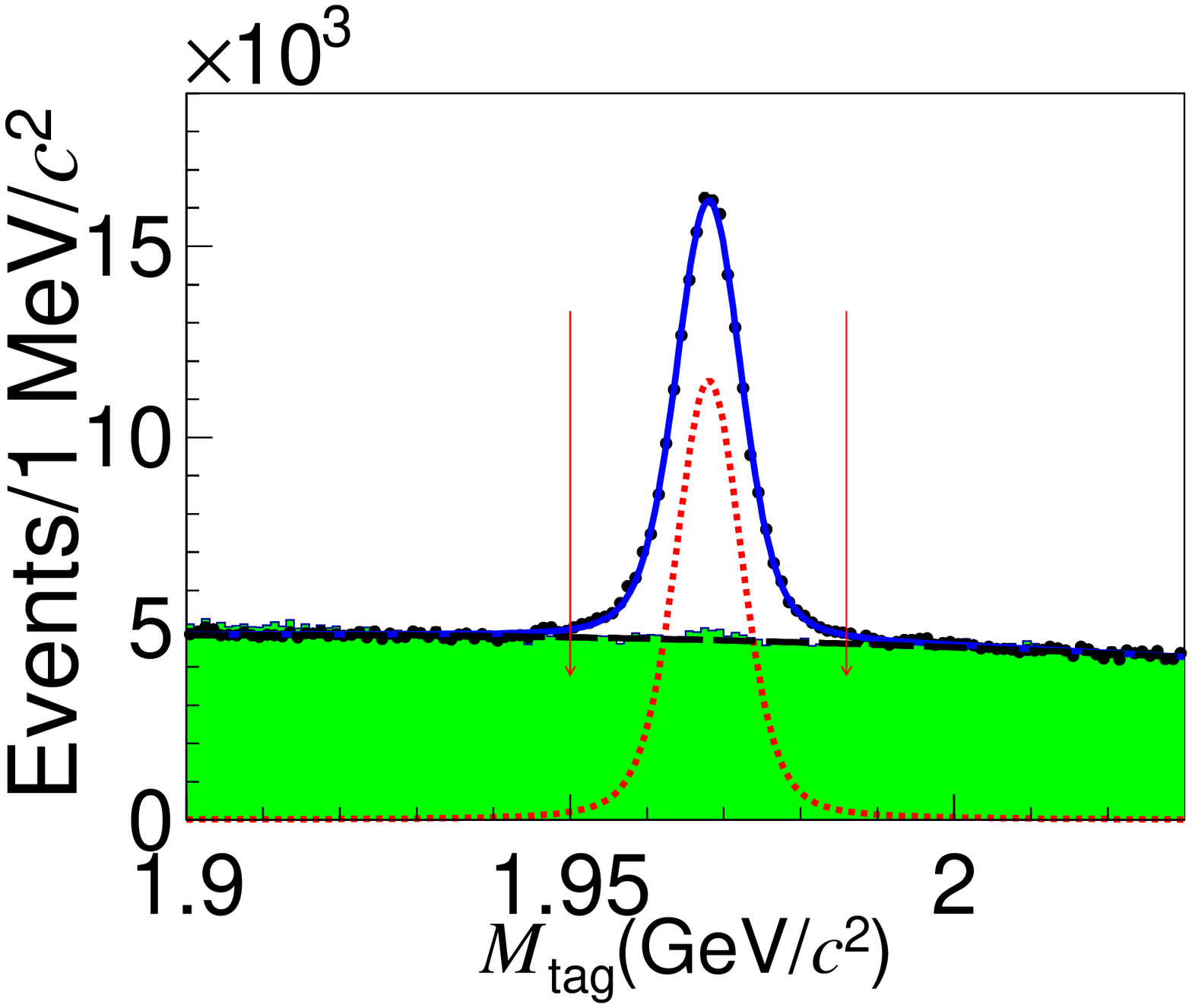}
\put(-20,65){(b)}
\end{minipage}
\caption{Fits to the $M_{{\rm tag}}$ spectra of (a) $D_{s}^{+} \rightarrow K_{S}^{0} K^{+}$
and (b) $D_{s}^{+} \rightarrow K^{-} K^{+} \pi^{+}$. 
The dots with error bars are data, the solid lines are the total fits, 
the dashed lines and the dotted lines are the shapes of signal and fitted background, respectively. 
The (green) filled histograms are the MC-simulated backgrounds. 
The $D_{s}$ signal regions are between the arrows.}
\label{fig:single tag}
\end{center}
\end{figure}

The ST yields are extracted from fits to the invariant mass spectra ($M_{{\rm tag}}$) of the 
ST $D_{s}^{-}$ candidates, the results are shown in Fig.~\ref{fig:single tag}.
The signal shape is modeled as a double Gaussian, while the background is parameterized as
a second-order Chebychev polynomial. 
The signal regions are defined as: $[1.948,\,1.991]~$GeV$/c^{2}$ for 
$D_{s}^{-} \rightarrow K_{S}^{0} K^{-}$ and $[1.950,\,1.986]~$GeV$/c^{2}$ for 
$D_{s}^{-} \rightarrow K^{+} K^{-} \pi^{-}$, respectively.
The ST yields determined by the fit for $D_{s}^{-} \rightarrow K_{S}^{0} K^{-}$ and 
$D_{s}^{-} \rightarrow K^{+} K^{-} \pi^{-}$ are $32751\pm313$ and $131862\pm773$, respectively.
For the tag mode $D_{s}^{-} \rightarrow K^{0}_{S} K^{-}$, a small peak in the background is observed 
in the signal region; this is due to $D^{-} \rightarrow K^{0}_{S} \pi^{-}$ events 
with the $\pi^{-}$ misidentified as a $K^{-}$.
From the generic MC sample, the yield of the $D^{-} \rightarrow K^{0}_{S} \pi^{-}$ background 
is estimated to be around 250, corresponding to about 0.2\% of the total ST yields, which is considered in the systematic uncertainty.
For the tag mode $D_{s}^{-} \rightarrow K^{+} K^{-} \pi^{-}$, a much smaller bump can also be found and 
the effect is negligible.

The DT events are reconstructed $D_{s}^{+}D_{s}^{-}$ pairs 
with $D^{-}_{s}$ reconstructed in a tag mode combined with 
$D_{s}^{+} \rightarrow \omega \pi^{+}$ or 
$D_{s}^{+} \rightarrow \omega K^{+}$ candidates, 
in which the $\omega$ is reconstructed in the $\pi^{+}\pi^{-}\pi^{0}$ final state. 
For the two $D_{s}$ candidates, we require that at least one of them has
$M_{{\rm rec}}$ greater than 2.10~GeV$/c^{2}$.
If there is more than one $D_{s}^{+}D_{s}^{-}$ pair candidate, 
the one with an average invariant mass of the two $D_{s}$ mesons 
closest to $m_{D_{s}}$ is chosen. 

For $D_{s}^{+} \rightarrow \omega K^{+}$, the background from the decay 
$D_{s}^{+} \rightarrow K_{S}^{0} K^{+} \pi^{0}$ is identical to the signal in the $M_{\rm rec}$ distribution and forms a peak around the $K^{*}(892)$ mass 
in the $\pi^{+}\pi^{-}\pi^{0}$ invariant mass ($M_{\pi^{+}\pi^{-}\pi^{0}}$) spectrum. 
Consequently, we further perform a $K_{S}^{0}$ veto to suppress this background. 
If the invariant mass of the $\pi^{+}\pi^{-}$ ($M_{\pi\pi}$) combination 
in $D_{s}^{+} \rightarrow \omega K^{+}$ 
signal candidate satisfies $|M_{\pi\pi}-m_{K_{S}^{0}}|<0.03$ GeV$/c^{2}$ and
the distance between the decay point and the IP has a significance of more than two 
standard deviations, the events are vetoed. 
This veto eliminates about 78\% of $D_{s}^{+} \rightarrow K_{S}^{0} K^{+} \pi^{0}$ 
background, while retaining about 97\% of signal events. 
After the $K_{S}^{0}$ veto, this background is found to be negligible according to the generic MC. 

A two dimensional (2D) extended unbinned likelihood 
fit is performed to the $M_{\pi^{+}\pi^{-}\pi^{0}}$ and the signal $D_{s}$
invariant mass ($M_{{\rm sig}}$) distributions to extract the signal yield. 
For $D^{+}_{s} \rightarrow \omega \pi^{+}$ candidates,
there are two $\pi^{+}\pi^{-}\pi^{0}$ combinations formed in each event.
In the data sample, there are 5 events with both $\pi^{+}\pi^{-}\pi^{0}$ combinations
retained in the fit range of $[0.60,\,0.95]~$GeV$/c^{2}$, but there is no evidence that these events create a peak.
This effect is negligible in the fit.

The $D_{s}$ signal is described by the MC simulated signal shape
convolved with a Gaussian ($f^{{\rm peak}}_{D_{s}}$). Here, the mean and the resolution of the Gaussian are 
fixed at the values determined from the fit to the sample of 
$D_{s}^{+} \rightarrow \pi^{+}\pi^{-}\pi^{0} \pi^{+}(K^{+})$ in data. 
The $\omega$ signal is represented by a Breit-Wigner (BW) convolved with a Gaussian ($f^{{\rm peak}}_{\omega}$),
where the mass and width are fixed to the PDG values~\cite{PDG}. 
The resolution of the Gaussian is fixed at the value determined from
the sample of $e^{+}e^{-} \rightarrow K^{+} K^{-} \omega$.
The combinatorial background in $M_{{\rm sig}}$ and $M_{\pi^{+}\pi^{-}\pi^{0}}$ spectra are 
parameterized by second-order Chebychev polynomials ($f^{{\rm poly}}_{D_{s}}$ and $f^{{\rm poly}}_{\omega}$). 

\begin{figure}[htbp]
\footnotesize
\begin{center}
\begin{minipage}[b]{0.23\textwidth}
\epsfig{width=0.98\textwidth,clip=true,file=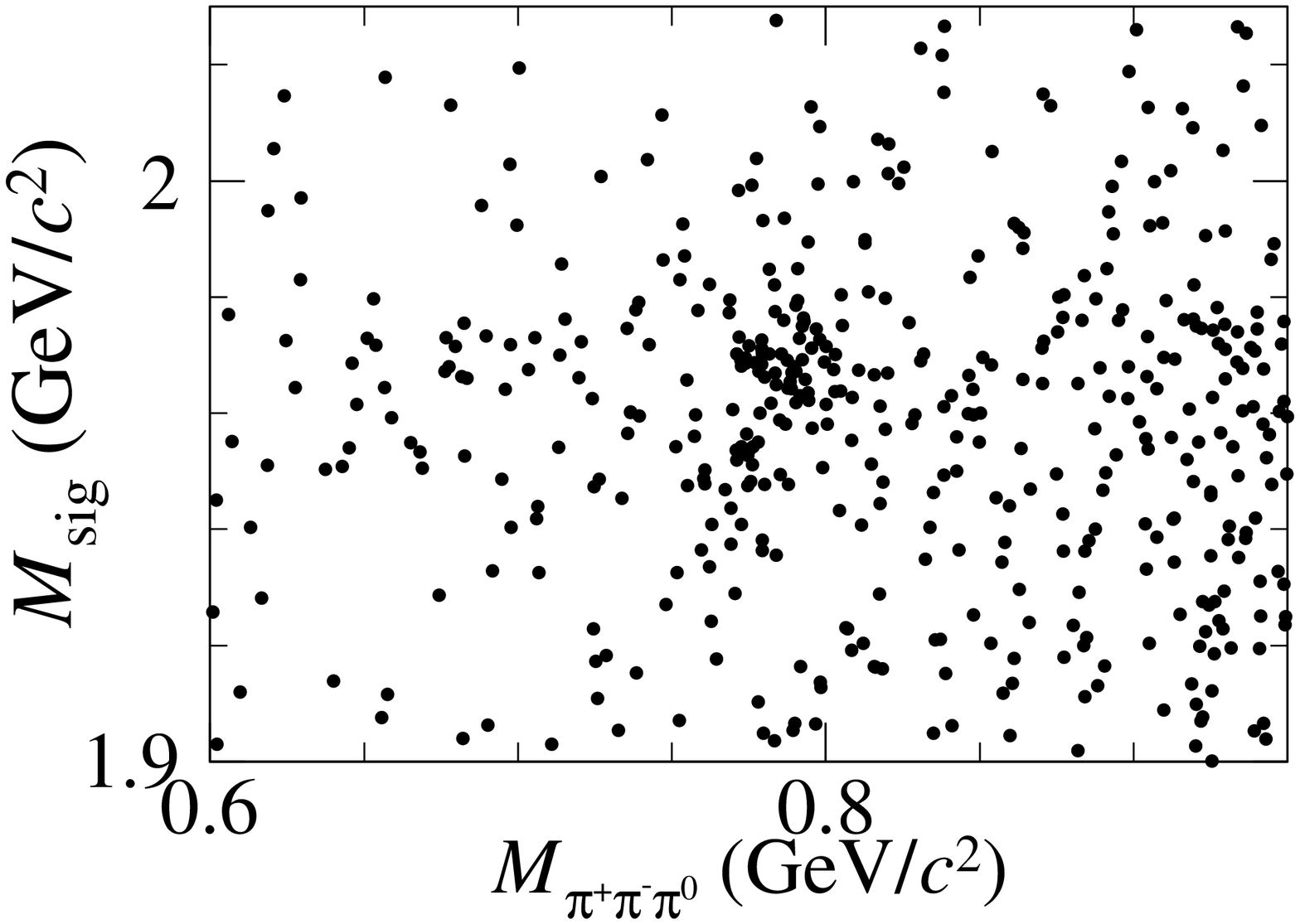}
\put(-80,72){(a)}
\end{minipage}
\begin{minipage}[b]{0.23\textwidth}
\epsfig{width=0.98\textwidth,clip=true,file=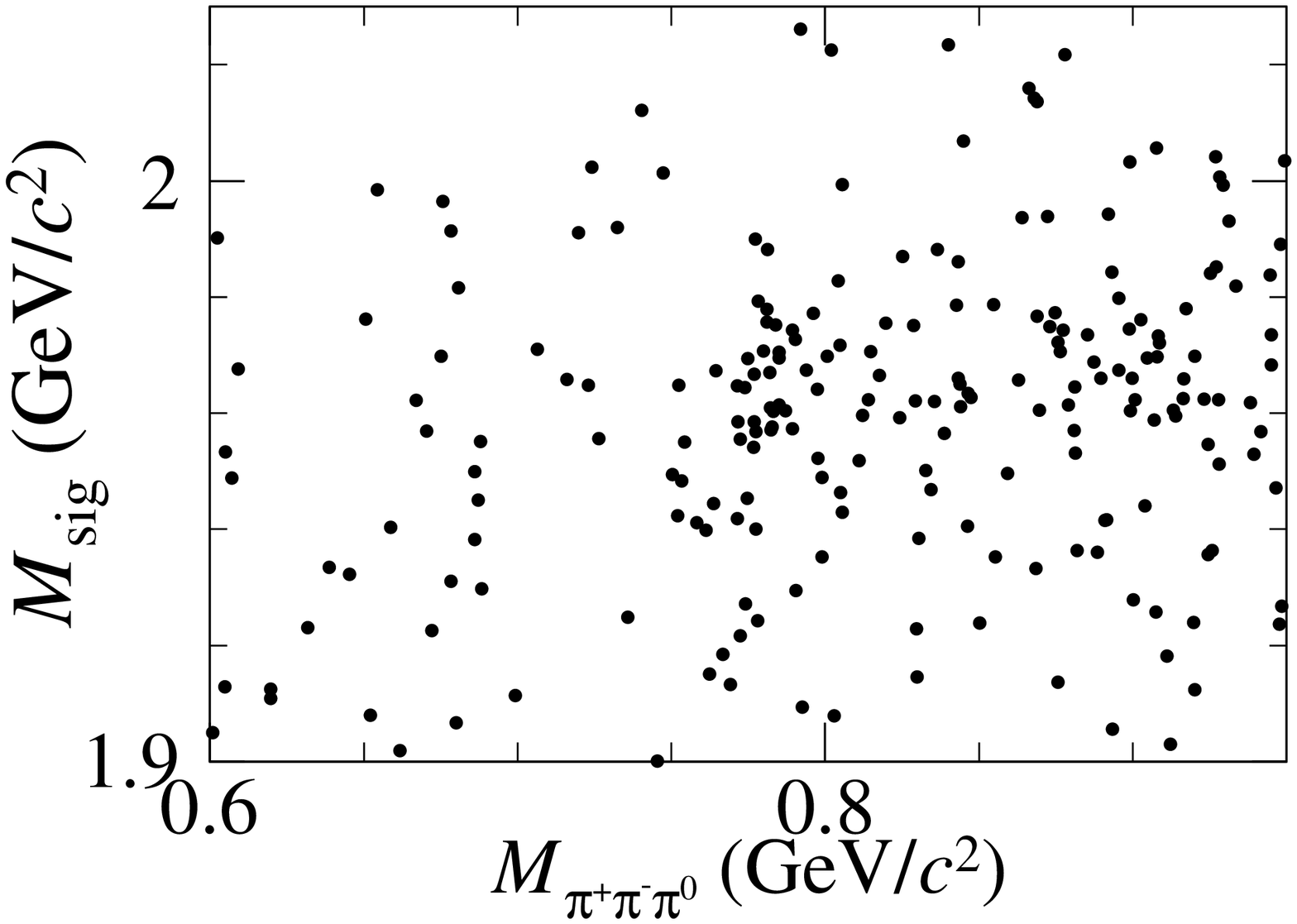}
\put(-80,72){(d)}
\end{minipage}
\begin{minipage}[b]{0.23\textwidth}
\epsfig{width=0.98\textwidth,clip=true,file=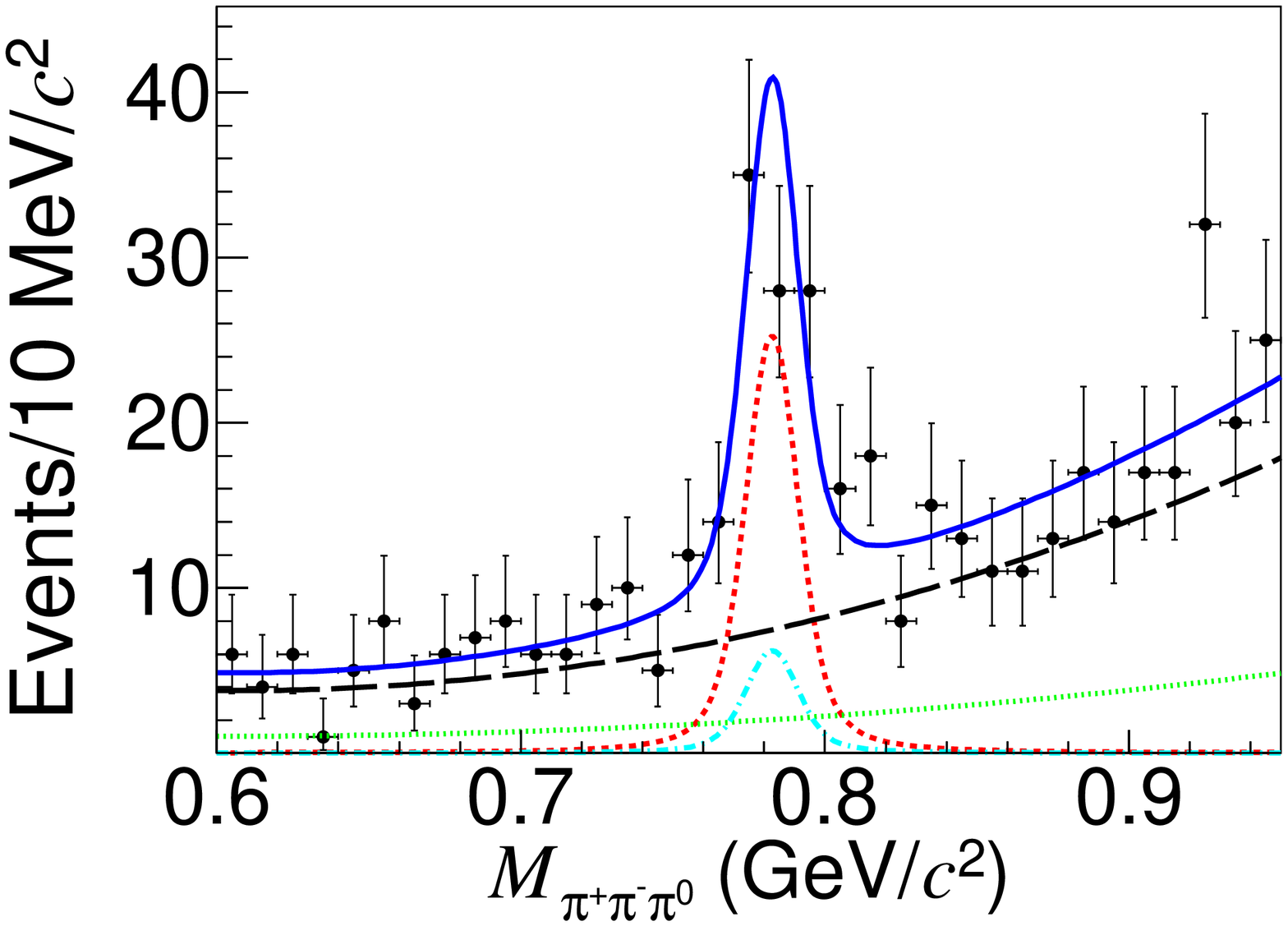}
\put(-80,72){(b)}
\end{minipage}
\begin{minipage}[b]{0.23\textwidth}
\epsfig{width=0.98\textwidth,clip=true,file=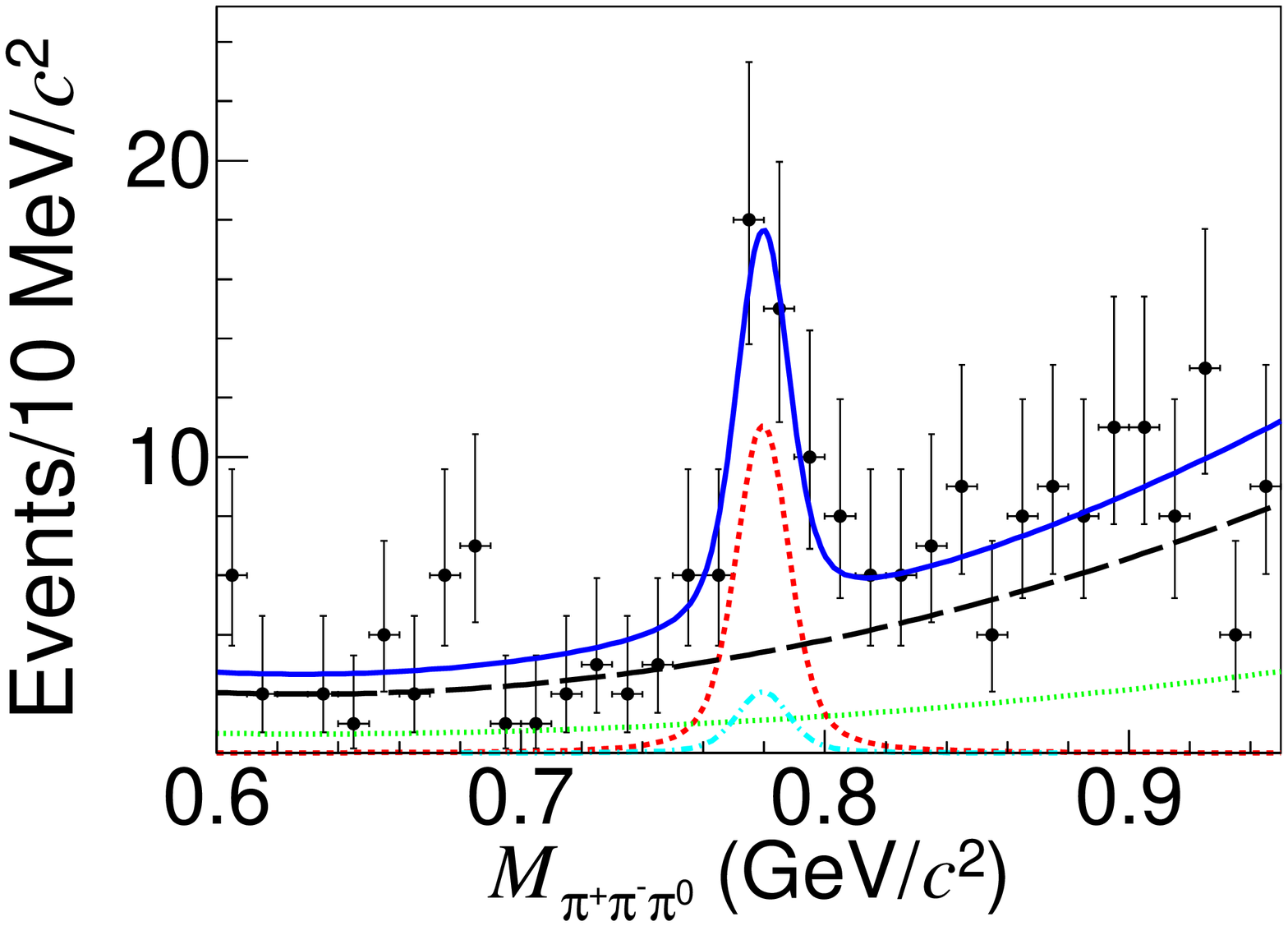}
\put(-80,72){(e)}
\end{minipage}
\begin{minipage}[b]{0.23\textwidth}
\epsfig{width=0.98\textwidth,clip=true,file=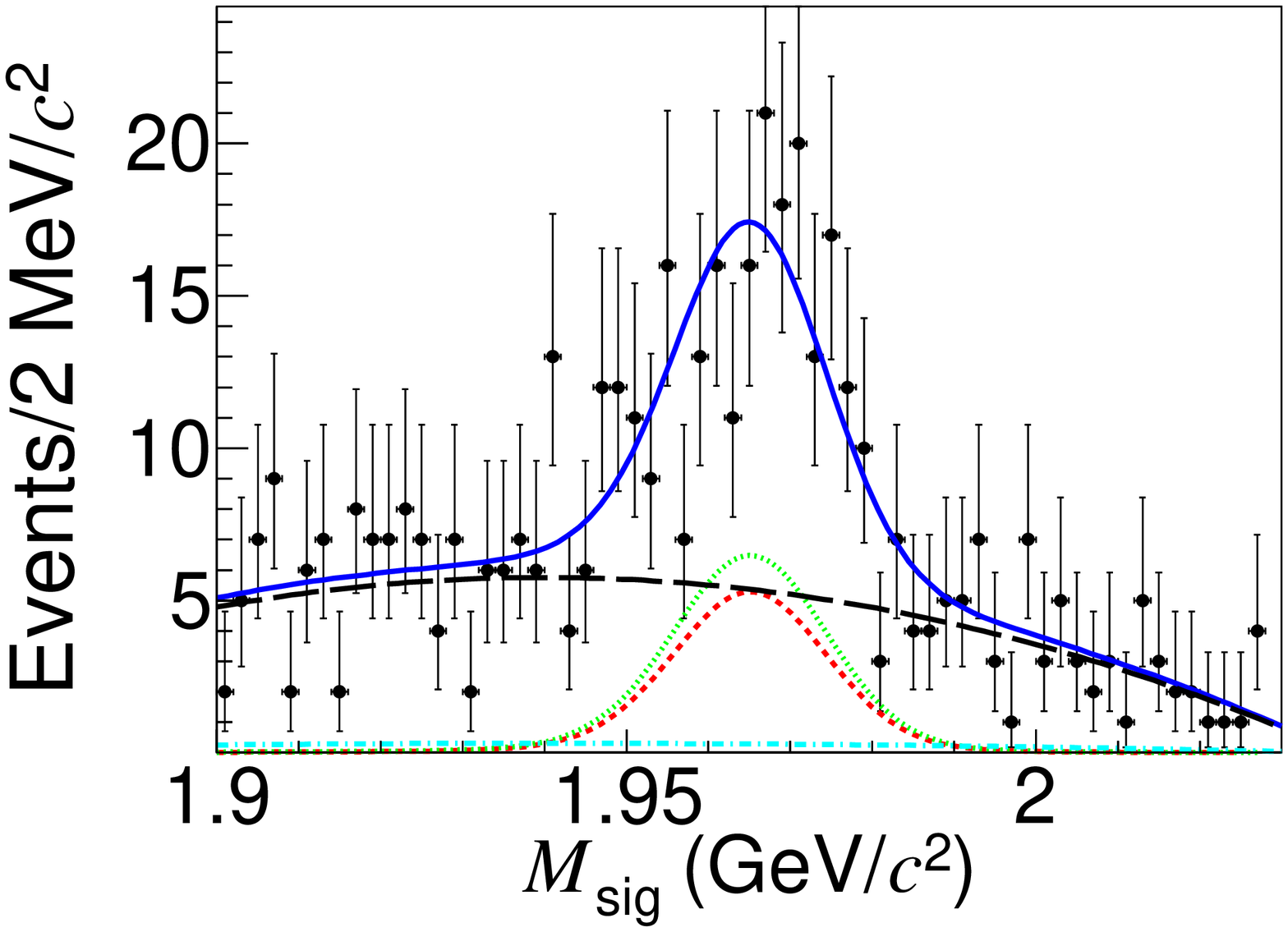}
\put(-80,72){(c)}
\end{minipage}
\begin{minipage}[b]{0.23\textwidth}
\epsfig{width=0.98\textwidth,clip=true,file=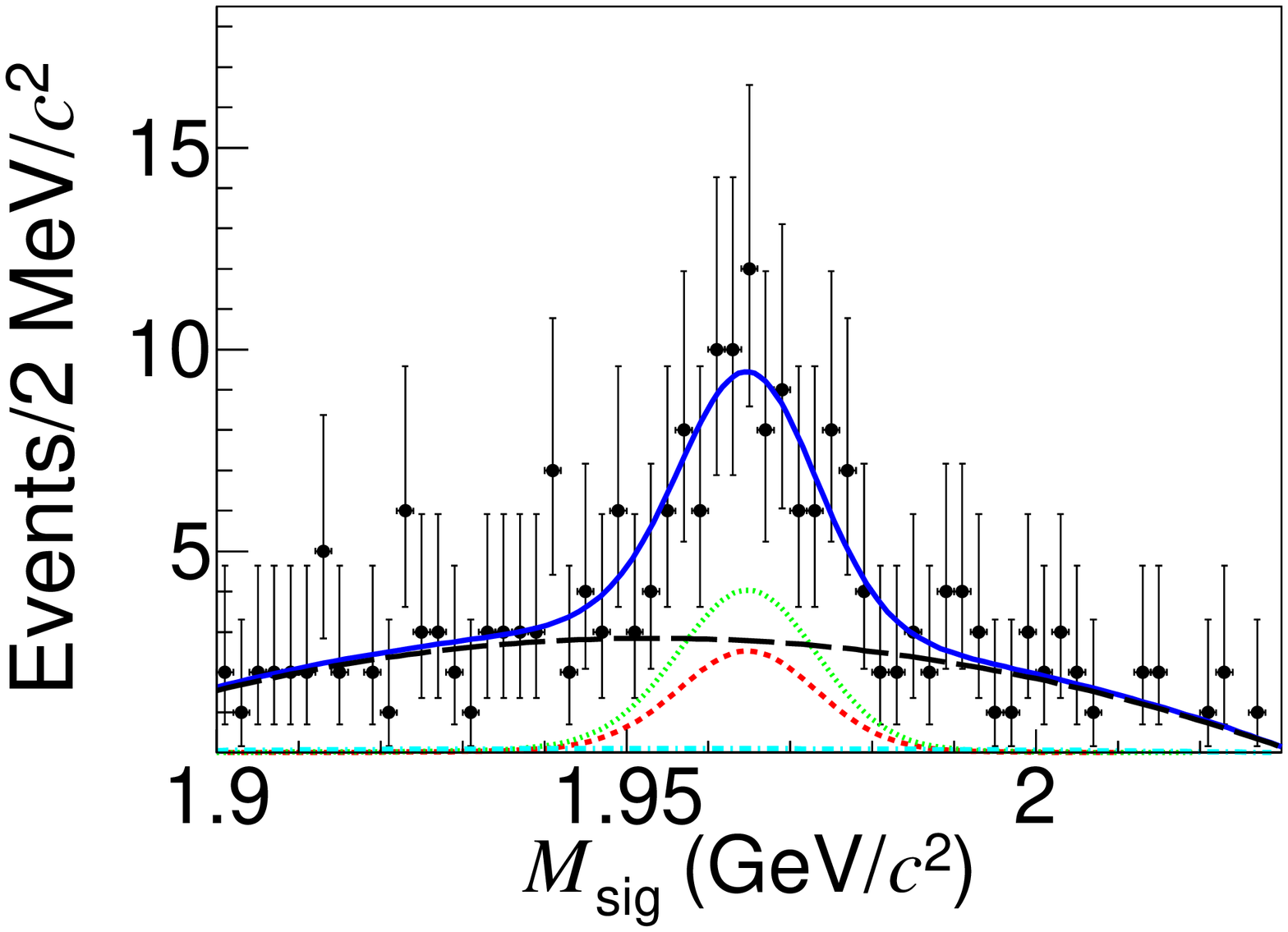}
\put(-80,72){(f)}
\end{minipage}
\caption{
The ((a) and (d)) scatter plots of $M_{{\rm sig}}$ versus $M_{\pi^{+}\pi^{-}\pi^{0}}$, 
fit results of ((b) and (e)) $M_{\pi^{+}\pi^{-}\pi^{0}}$, 
and fit results of ((c) and (f)) $M_{{\rm sig}}$ 
for (a-c) $D^{+}_{s} \rightarrow \omega \pi^{+}$ and (d-f)
$D^{+}_{s} \rightarrow \omega K^{+}$. 
In the fits, the dots with error bars are data, the (blue) solid lines describe the total fits, 
the (red) dashed lines describe the signal shape and the (green) dotted, (cyan) dash-dotted, 
and (black) long dashed lines describe the BKGI, BKGII, and BKGIII, respectively.} 
\label{fig:result}
\end{center}
\end{figure}

The scatter plots of $M_{{\rm sig}}$ versus $M_{\pi^{+}\pi^{-}\pi^{0}}$ for the two signal decays are 
shown in Figs.~\ref{fig:result}(a,b), from which no obvious correlation 
between $M_{{\rm sig}}$ and $M_{\pi^{+}\pi^{-}\pi^{0}}$ is found, which is also confirmed by signal MC. 
The 2D fit model is then constructed as following. 
The signal shape is modeled as the product of $f^{{\rm peak}}_{D_{s}}$ and $f^{{\rm peak}}_{\omega}$. 
The background that does not peak in both $M_{\pi^{+}\pi^{-}\pi^{0}}$ and $M_{{\rm sig}}$ distributions (BKGI)
is modeled as the product of $f^{{\rm poly}}_{D_{s}}$ and $f^{{\rm poly}}_{\omega}$. 
The background with an $\omega$ that peaks in the $M_{\pi^{+}\pi^{-}\pi^{0}}$ distribution (BKGII)
is modeled as the product of $f^{{\rm poly}}_{D_{s}}$ and $f^{{\rm peak}}_{\omega}$. 
The background from $D_{s}^{+} \rightarrow \pi^{+}\pi^{-}\pi^{0}\pi^{+}(K^{+})$ 
that only has a peak in the $M_{{\rm sig}}$ distribution (BKGIII)  
is modeled as the product of $f^{{\rm peak}}_{D_{s}}$ and $f^{{\rm poly}}_{\omega}$. 
The parameters in $f^{{\rm poly}}_{D_{s}}$ and $f^{{\rm poly}}_{\omega}$ obtained in the 2D fit are 
consistent with the results obtained in the individual fits to the $M_{{\rm sig}}$ and $M_{\pi^{+}\pi^{-}\pi^{0}}$ spectra, respectively. 
Therefore, in the 2D fit, yields of signal and backgrounds are 
determined by the fit and the other parameters are fixed at the values from the individual fits. 
From the 2D fits, we obtain 
$65.0\pm11.6$ $D^{+}_{s} \rightarrow \omega \pi^{+}$ signals 
and $28.5\pm7.8$ $D^{+}_{s} \rightarrow \omega K^{+}$ signals with statistical  
significances of 6.7$\sigma$ and 4.4$\sigma$, respectively.
The fit results are shown in Figs.~\ref{fig:result}(b,c,e,f). 

The ST and DT efficiencies are determined from the generic MC and 
signal MC samples, respectively. All efficiencies are summarized in Table~\ref{tab:tag}.
\begin{table}[htp]
\begin{center}
\caption{The ST efficiencies $\epsilon_{{\rm tag}}$ and DT efficiencies $\epsilon_{{\rm tag,sig}}$.}
\begin{tabular}{cccc}\hline
Tag mode                                              & $\epsilon_{{\rm tag}}$ (\%) &$\epsilon_{{\rm tag},\omega \pi^{+}}$ (\%) &$\epsilon_{{\rm tag},\omega K^{+}}$ (\%)    \\ \hline
$D^{-}_{s} \rightarrow K_{S}^{0}(\pi^{+}\pi^{-})K^{-}$& $51.38\pm0.25$              &$12.53\pm0.13$                         & $10.74\pm0.11$\\
$D^{-}_{s} \rightarrow K^{+}K^{-}\pi^{-}$             & $38.44\pm0.08$              &$9.79\pm0.06$                          & $8.81\pm0.06$\\
\hline
\end{tabular}
\label{tab:tag}
\end{center}
\end{table}
Using Eq.~(\ref{Eq:Br}) and 
the world averaged BFs of $\omega \rightarrow \pi^{+}\pi^{-}\pi^{0}$ 
and $\pi^{0} \rightarrow \gamma\gamma$~\cite{PDG},  
the BFs are measured to be: 
$\mathcal{B}(D^{+}_{s} \rightarrow \omega \pi^{+}) = (1.77\pm0.32) \times 10^{-3}$ and
$\mathcal{B}(D^{+}_{s} \rightarrow \omega K^{+}) = (0.87\pm0.24) \times 10^{-3}$, where the uncertainties are statistical.

The systematic uncertainties are investigated and  
are summarized in Table~\ref{tab:sys_unc}.
For each decay, the total systematic uncertainty is 
obtained by adding the individual terms in quadrature.

The uncertainties due to the $M_{{\rm rec}}$ requirement and momentum requirement 
on pion are estimated with the control sample of $D_{s}^{+} \rightarrow \pi^{+}\pi^{-}\pi^{+}\eta$, with $\eta^{\prime}$ 
decays removed by requiring the invariant mass of $\pi^{+}\pi^{-}\eta$ to be greater than 1.0 GeV$/c^{2}$. 
The uncertainty due to the $K_{S}^{0}$ veto is estimated with the control sample of $D^{0} \rightarrow K_{S}^{0}\omega$. 
The differences between the efficiencies caused by corresponding selection criterion obtained from data and MC simulation are taken as the systematic uncertainties.

The uncertainties for charged tracks selection are 
determined to be 0.5\%/track for PID and 1.0\%/track for tracking using the control sample of 
$e^{+}e^{-} \rightarrow K^{+}K^{-}\pi^{+}\pi^{-}$. 
The uncertainty of the $\pi^{0}$ reconstruction efficiency is investigated with the control sample of $e^{+}e^{-}\rightarrow K^{+}K^{-}\pi^{+}\pi^{-}\pi^{0}$, 
and is determined to be 1.8\% (1.9\%) for $D_{s}^{+} \rightarrow \omega \pi^{+} (K^{+})$. The uncertainty due to the MC statistics is 0.6\%.

The uncertainty due to the signal shape is estimated by varying the masses and resolutions of the $\omega$ and $D_{s}$ within their uncertainties.
The uncertainty due to the background shape is investigated by narrowing the fit ranges of $M_{\pi^{+}\pi^{-}\pi^{0}}$ and $M_{{\rm sig}}$ to 
$[0.65,\,0.90]~$GeV$/c^{2}$ and $[1.91,\,2.02]~$GeV$/c^{2}$, respectively, and replacing the second-order Chebychev polynomial 
in $f^{{\rm poly}}_{D_{s}}$ and $f^{{\rm poly}}_{\omega}$ by a first-order Chebychev polynomial.  
The uncertainty related to ST yields is estimated using the MC-simulated shape to replace the Gaussian function in the ST yields determination. 
Here, the effect from the bump under the $D^{-}_{s} \rightarrow K_{S}^{0}K^{-}$ signal region is also taken into account.  
For each alteration of the fit configuration the measurements are re-performed. 
The largest change to the BF is taken as the corresponding uncertainty.
The uncertainty from the shape and fit ranges effect of background in ST yields is found to be negligible.
The uncertainty due to the fit procedure is investigated 
by studying ten statistically independent samples of generic MC events with the same size as data.
With the same method as used in data analysis, the average measured BF is found to have a relative difference of 0.8\% with respect to the input value. 
This difference is taken as the uncertainty from the fit procedure.
The uncertainty related to the assumed BFs for $\omega \rightarrow \pi^{+}\pi^{-}\pi^{0}$
and $\pi^{0}\rightarrow \gamma\gamma$ is taken from the PDG~\cite{PDG}.

\begin{table}[htp]
\begin{center}
\caption{Relative systematic uncertainties (\%) in the BF measurements.}
\begin{tabular}{c|cc} \hline
Source & $D^{+}_{s} \rightarrow \omega \pi^{+}$ & $D^{+}_{s} \rightarrow \omega K^{+}$\\
\hline
$M_{{\rm rec}}$ requirement                   & \multicolumn{2}{c}{0.1} \\ 
Momentum requirement on pion                  & \multicolumn{2}{c}{1.7} \\ 
$K_{S}^{0}$ veto                              & -  &  0.1  \\  
PID of $K^{\pm}$,$\pi^{\pm}$& 1.5 & 1.5 \\ 
Tracking of $K^{\pm}$,$\pi^{\pm}$& 3.0 & 3.0 \\ 
$\pi^{0}$ reconstruction& 1.8 & 1.9 \\ 
MC statistics           & 0.6 & 0.6 \\ 
Background shape& 4.1 & 4.6 \\ 
Signal shape    & 3.3 & 5.3 \\ 
ST yield        & 1.3 & 1.3 \\ 
Fit procedure    & \multicolumn{2}{c}{0.8} \\  
$\mathcal{B}(\omega \rightarrow \pi^{+}\pi^{-}\pi^{0})$ \& $\mathcal{B}(\pi^{0}\rightarrow \gamma\gamma)$& \multicolumn{2}{c}{0.8} \\
\hline
Total       & 7.0 & 8.4 \\ 
\hline
\end{tabular}
\label{tab:sys_unc}
\end{center}
\end{table}

In summary, we observe the $W$-annihilation decay $D^{+}_{s} \rightarrow \omega \pi^{+}$ with a significance of 6.7$\sigma$
and measure its BF to be
$(1.77\pm0.32_{{\rm stat.}}\pm0.12_{{\rm sys.}}) \times 10^{-3}$. 
This measurement provides critical information to determine the non-perturbative $W$-annihilation amplitudes. 
The significantly improved precision benefits the investigations of the underlying dynamics in charmed hadronic decays, and will allow better 
predictions for the BFs and direct CPV of decays involving $W$-annihilation~\cite{Cheng:2012wr,Beast1,HaiYangCheng1,HaiYangCheng2}. 
Among these decays, $D^{+}_{s} \rightarrow \omega K^{+}$ is interest for its possibly large CPV. 
We find the first evidence for this decay with a significance of 4.4$\sigma$. 
Its BF is measured to be
$(0.87\pm0.24_{{\rm stat.}}\pm0.07_{{\rm sys.}}) \times 10^{-3}$. 
According to Ref.~\cite{Beast1}, our result implies that the $\rho-\omega$ mixing is negligible 
and direct CP asymmetry is expected at the level of $-0.6 \times 10^{-3}$. 

The authors wish to thank Fusheng Yu for useful discussions. 
The BESIII collaboration thanks the staff of BEPCII and the IHEP 
computing center for their strong support. 
This work is supported in part by National Key Basic Research Program of China 
under Contract No. 2015CB856700; National Natural Science Foundation of China (NSFC) 
under Contracts Nos. 11425524, 11625523, 11635010; 
the Chinese Academy of Sciences (CAS) Large-Scale Scientific Facility Program; 
the CAS Center for Excellence in Particle Physics (CCEPP); 
Joint Large-Scale Scientific Facility Funds of the NSFC and CAS 
under Contracts Nos. U1332201, U1532257, U1532258; 
CAS Key Research Program of Frontier Sciences under Contracts 
Nos. QYZDJ-SSW-SLH003, QYZDJ-SSW-SLH040; 
100 Talents Program of CAS; National 1000 Talents Program of China; 
INPAC and Shanghai Key Laboratory for Particle Physics and Cosmology; 
German Research Foundation DFG under Contracts Nos. Collaborative Research Center CRC 1044, FOR 2359; 
Istituto Nazionale di Fisica Nucleare, Italy; 
Koninklijke Nederlandse Akademie van Wetenschappen (KNAW) under Contract No. 530-4CDP03; 
Ministry of Development of Turkey under Contract No. DPT2006K-120470; 
National Science and Technology fund; 
The Swedish Research Council; 
U. S. Department of Energy under Contracts Nos. DE-FG02-05ER41374, DE-SC-0010118, DE-SC-0010504, DE-SC-0012069; University of Groningen (RuG) and the Helmholtzzentrum fuer Schwerionenforschung GmbH (GSI), Darmstadt; 
WCU Program of National Research Foundation of Korea under Contract No. R32-2008-000-10155-0.



\begin{thebibliography}{99}

%
%
%
\bibitem{Cheng:2012wr} H.~Y.~Cheng and C.~W.~Chiang, Phys.\ Rev.\ D {\bf 85}, 034036 (2012); 
                       Erratum: [Phys.\ Rev.\ D {\bf 85}, 079903 (2012)]. 

\bibitem{Li:2012cfa} H.~n.~Li, C.~D.~Lu and F.~S.~Yu, Phys.\ Rev.\ D {\bf 86}, 036012 (2012). 

\bibitem{Beast1} Q. Qin, H. N. Li, C. D. L\"u and F. S. Yu, Phys. Rev. D {\bf 89}, 054006 (2014).



\bibitem{HaiYangCheng1} H. Y. Cheng and C. W. Chiang, Phys. Rev. D {\bf 81}, 074021 (2010).


\bibitem{HaiYangCheng2} H. Y. Cheng, C. W. Chiang and A. L. Kuo, Phys. Rev. D {\bf 93}, 114010 (2016).
 
\bibitem{CLEOobserve} R. Balest {\it et al}. (CLEO Collaboration), Phys. Rev. Lett. {\bf 79}, 1436 (1997).


\bibitem{CLEOcomega} J. Y. Ge {\it et al}. (CLEO Collaboration), Phys. Rev. D {\bf 80}, 051102(R) (2009).
%
\bibitem{tagmethod} R. M. Baltrusaitis {\it et al}. (Mark III Collaboration), Phys. Rev. Lett. {\bf 56}, 2140 (1986).

\bibitem{detector} M. Ablikim {\it et al}. (BESIII Collaboration), Nucl. Instrum. Methods Phys. Res., Sect. A {\bf 614}, 345 (2010).

\bibitem{MRPC} X.~Wang {\it et al.}, JINST {\bf 11}, no. 08, C08009 (2016).

\bibitem{sim} S. Agostinelli {\it et al}. (GEANT4 Collaboration), Nucl. Instrum. Methods Phys. Res., Sect. A {\bf 506}, 250 (2003).

%
%
%
%

\bibitem{Ping:2013jka} R.~G.~Ping, Chin.\ Phys.\ C {\bf 38}, 083001 (2014).

\bibitem{EvtGen} D. J. Lange, Nucl. Instrum. Methods Phys. Res., Sect. A {\bf 462}, 152 (2001); 
                 R. G. Ping, Chin. Phys. C {\bf 32}, 599 (2008).

\bibitem{KKMC} S. Jadach, B. F. L. Ward and Z. Was, Phys. Rev. D {\bf 63}, 113009 (2001).

\bibitem{BABAYAGA}C. M. Carloni Calame {\it et al}., Nucl. Phys. B (Proc. Suppl.) {\bf 131}, 48 (2004).


\bibitem{PDG} M. Tanabashi {\it et al.} (Particle Data Group), Phys. Rev. D 98, 030001 (2018).

\bibitem{QinXiaoshuai} M. Ablikim {\it et al}. (BESIII Collaboration), Phys. Rev. Lett. {\bf 116}, 082001 (2016).

\end{thebibliography}
\end{document}